\documentclass[aps, pra, reprint, article]{revtex4-1}

\usepackage{amsmath,amsfonts,amssymb}						
\usepackage{braket} 										
\usepackage{nicefrac} 										
\usepackage{mhsetup} 										
\usepackage{mathtools}										
\usepackage[version=3]{mhchem} 								
\usepackage{siunitx} 										

\DeclareSIUnit\kay{\per \cm}

\usepackage{xcolor}

\usepackage{epstopdf}		

\usepackage{multirow}
\usepackage{diagbox}										

\newcommand*{\tabref}[1]{\tablename~\ref{#1}}
\newcommand*{\figref}[1]{\figurename~\ref{#1}}
\newcommand*{\equref}[1]{Eq.~\eqref{#1}}

\newcommand{\bands}[2]{\mbox{(#1--#2)}}		

\newcommand{\loS}{^{\prime\prime}}
\newcommand{\upS}{^{\prime}}

\newcommand{\fone}{$\mathrm{F}_1$}
\newcommand{\ftwo}{$\mathrm{F}_2$}

\newcommand{\Xstate}{$\mathrm{X}(1)^2\Sigma^+$}
\newcommand{\bstate}{$(2)^2\Sigma^+$}
\newcommand{\astate}{$(1)^2\Pi$}
\newcommand{\KCa}{\ce{^{39}K^{40}Ca}~}

\begin{document}

\title[\ce{KCa} \Xstate \ and \bstate \  states]{Laser and Fourier transform spectroscopy of \ce{KCa}}

\date{\today}

\begin{abstract}
	\ce{KCa} was produced in a heatpipe oven and its thermal emission spectrum around \SI{8900}{\kay} was recorded by a high resolution Fourier transform spectrometer. In addition, many selected transitions of this spectrum between deeply bound vibrational levels of the \Xstate \ and \bstate \  states were studied using laser excitation to facilitate the assignment of the lines. The ground state is described for $v^{\prime\prime} =$ 0 -- 5 and the \bstate \  state for $v^{\prime} =0$ -- 8  with rotational levels up to 175. For both states, Dunham coefficients, spin-rotation parameters and potential energy curves are derived.
\end{abstract}

\author{Julia Gerschmann}
\affiliation{ Leibniz Universit\"at Hannover, Institute of Quantum Optics, Welfengarten 1, 30167 Hannover, Germany}
\affiliation{Leibniz Universit\"at Hannover, Laboratory for Nano- and Quantum Engineering,  Schneiderberg 39, 30167 Hannover, Germany}
\email{gerschmann@iqo.uni-hannover.de}

\author{Erik Schwanke}
\affiliation{ Leibniz Universit\"at Hannover, Institute of Quantum Optics, Welfengarten 1, 30167 Hannover, Germany}
\affiliation{Leibniz Universit\"at Hannover, Laboratory for Nano- and Quantum Engineering,  Schneiderberg 39, 30167 Hannover, Germany}

\author{Asen Pashov}
\affiliation{ Department of Physics, Sofia University, 5 James Bourchier Boulevard, 1164 Sofia, Bulgaria}

\author{Horst Kn\"ockel}
\affiliation{ Leibniz Universit\"at Hannover, Institute of Quantum Optics, Welfengarten 1, 30167 Hannover, Germany}
\affiliation{Leibniz Universit\"at Hannover, Laboratory for Nano- and Quantum Engineering,  Schneiderberg 39, 30167 Hannover, Germany}

\author{Silke Ospelkaus}
\affiliation{ Leibniz Universit\"at Hannover, Institute of Quantum Optics, Welfengarten 1, 30167 Hannover, Germany}
\affiliation{Leibniz Universit\"at Hannover, Laboratory for Nano- and Quantum Engineering,  Schneiderberg 39, 30167 Hannover, Germany}

\author{Eberhard Tiemann}
\affiliation{ Leibniz Universit\"at Hannover, Institute of Quantum Optics, Welfengarten 1, 30167 Hannover, Germany}
\affiliation{Leibniz Universit\"at Hannover, Laboratory for Nano- and Quantum Engineering,  Schneiderberg 39, 30167 Hannover, Germany}

%

\maketitle
	
\section*{introduction}

	Molecules consisting of one alkali-metal atom and one alkaline-earth atom receive rising interest for their prospective application in the field of ultracold quantum gases because their ground state  \Xstate \ with its electric and magnetic dipole moments offers advantageous properties (see e.g. references \cite{tscherbul_controlling_2006,pasquiou_quantum_2013,roy_photoassociative_2016}) for testing and/or demonstrating fundamental properties of quantum gases.
	After investigating \ce{LiCa}\cite{stein_spectroscopic_2013} and \ce{LiSr}\cite{LiSrFTS}, we proceeded to the heavier \ce{KCa} for opening perspectives for spectroscopic studies of even heavier species like \ce{RbSr} for which already ultracold ensembles of Rb and Sr were prepared\cite{pasquiou_quantum_2013}. Compared to LiCa and LiSr, however, the experimental study becomes more difficult because of the higher density of spectral lines of  \ce{KCa} and the always accompanying molecule \ce{K_2}. We present a spectroscopic observation of \ce{KCa} and its first rovibrational analysis from which the bottoms of the potential energy curves of the ground state \Xstate \ and the first excited state \bstate \  are derived.
	
\begin{figure}
		\includegraphics[width = \columnwidth]{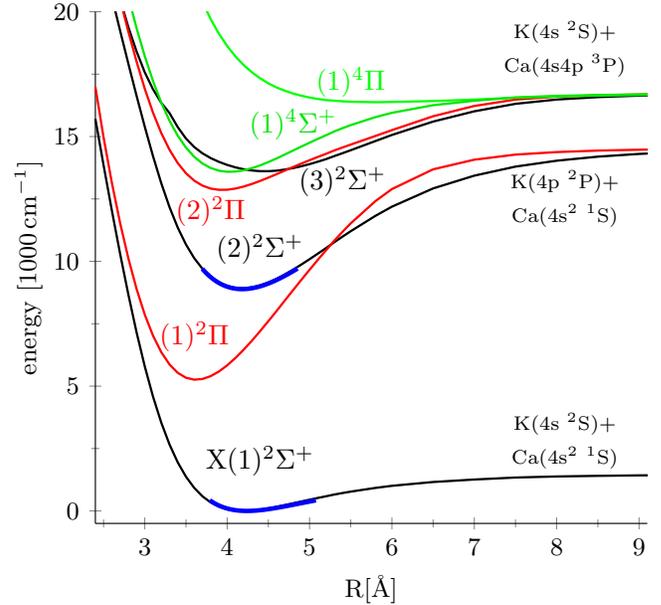}
		\caption{ (Colour online) Potential energy curves of \ce{KCa} from ab initio calculations\cite{pototschnig_vibronic_2017}. Thick curves represent PECs derived form the experimental data of this work.}
		\label{fig:PotentialCurves}
	\end{figure}

	Ab initio calculations for \ce{KCa} have been performed by other groups\cite{gopakumar_dipole_2014, pototschnig_vibronic_2017}. \figref{fig:PotentialCurves} shows a part of the potential energy scheme of \ce{KCa} for the lowest atom pair asymptote K(4s $^2$S) + Ca(4s$^2$ $^1$S), and the excited asymptote K(4p $^2$P) + Ca(4s$^2$ $^1$S) and K(4s $^2$S) + Ca(4s4p $^3$P).
	We started with recording the thermal emission spectrum of \ce{KCa} in a heatpipe to locate the general position of the spectrum and to discriminate it from the \ce{K_2} spectrum.
	 The observed spectrum was found in the near infrared region, expected from the ab initio results shown in \figref{fig:PotentialCurves}. Laser excitations of the molecule were performed and were essential for an unambiguous assignment of the dense and overlapping spectral structure. Methods to interpret the results of such experiments for gathering structural information about the molecule are described.
	For the observed \bstate \  -- \Xstate \ system, molecular parameters are derived. A comparison of  results from ab initio studies with experimental findings of this work is presented.

\section{experiment}
\label{sec:experiment}

\begin{figure}
    \centering
    \includegraphics[width=\columnwidth]{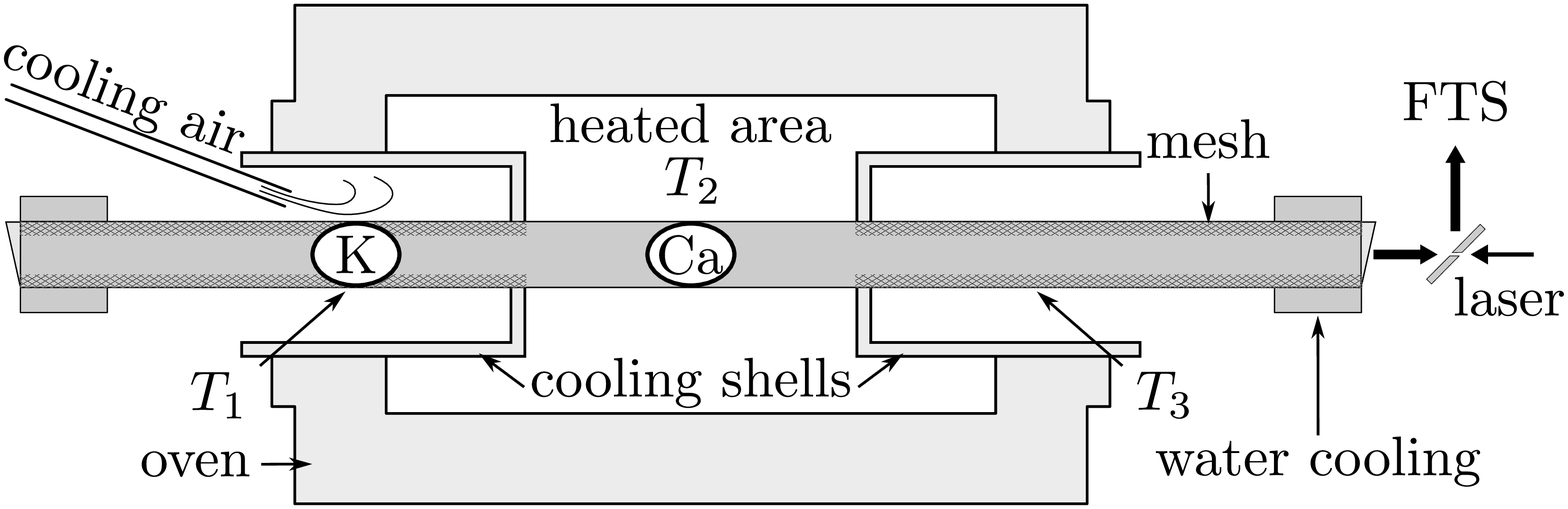}
    \caption[KCa heatpipe]{Experimental setup of the heatpipe. The heatpipe is driven by an oven. Three temperature ranges are created by cooling shells: $T_1$: 260-\SI{340}{\celsius}, $T_2$: \SI{820}{\celsius} and $T_3$: 450-\SI{650}{\celsius}.}
    \label{fig:kcahp}
\end{figure}

The spectroscopic setup consists  of a heatpipe oven for the molecular gas and an optical system for imaging the thermal emission spectrum into a Fourier Transform Spectrometer (FTS).
The heatpipe is shown in \figref{fig:kcahp}. It is a 88 cm long steel tube with a diameter of 3 cm. The middle section of the heatpipe is enclosed by an oven. The ends of the heatpipe are cooled to room temperature via water-cooling so that no metal vapour can reach the BK7 windows. The windows are tilted to avoid back reflections. The outer regions of the heatpipe are internally covered with a steel mesh of approximately 30 cm in length, so that the condensed metal can flow back into the heated middle section.  Additionally, shells cooled with forced airflow are installed at these regions (see \figref{fig:kcahp}) to create lower temperature areas and an appropriate temperature gradient for the very different vapour pressures of Ca and K. During the experimental run, crystals grow inside the heatpipe, producing laser stray light and finally blocking the optical path. That significantly reduces the measuring time to about five hours. Therefore, they must be melted intermittently by changing the heatpipe position in the oven. The use of cooling shells slows crystal growth.

To prepare the heatpipe, 10 g of calcium are placed in the middle of the heatpipe and melted under an atmosphere of 200-300 \si{\milli\bar} of argon at an oven temperature of up to \SI{1000}{\celsius}. After the heatpipe is cooled down, 5 g of potassium are placed on the mesh at a distance of about 25 cm from calcium on the outer region of the heatpipe not facing the spectrometer. Potassium is melted at the same buffer gas pressure as for melting calcium and an oven temperature of \SI{400}{\celsius}. Afterwards, the cooling shells are installed.

The light emerging from the heatpipe is imaged into the FTS (IFS 120HR, Bruker) via a mirror and a lens system and finally detected by an IR-enhanced silicon avalanche photodiode (S11519-30, Hamamatsu). The laser light is introduced counter-propagating the imaging path of the thermal emission to avoid direct laser radiation into the FTS. For this purpose, the mirror is slitted and the laser beam propagates along the optical axis within the heatpipe, indicated in \figref{fig:kcahp}.

To produce the KCa molecules, the region of the heatpipe at the calcium position is heated to approximately \SI{820}{\celsius}. By varying the air flow, the cooling shells create an region with 260-\SI{340}{\celsius} on the side of the potassium and an region with 450-\SI{650}{\celsius} on the other side. During the measurements, the pressure of the buffer gas, argon, is about 50 mbar. 

In addition to KCa, \ce{K_2} molecules are forming in the heatpipe even at low temperatures and dominate the gas spectrum. Since the observed emission lines of both molecules lie in different spectral ranges, this does not affect the observation of KCa. We use \ce{K_2} to align the imaging of the fluorescence region by exciting \ce{K_2} molecules with a HeNe-laser and adjusting the mirror so that only the fluorescence, but no stray light from the walls, is recorded.

\begin{figure}
	\centering
	\includegraphics[width=\columnwidth]{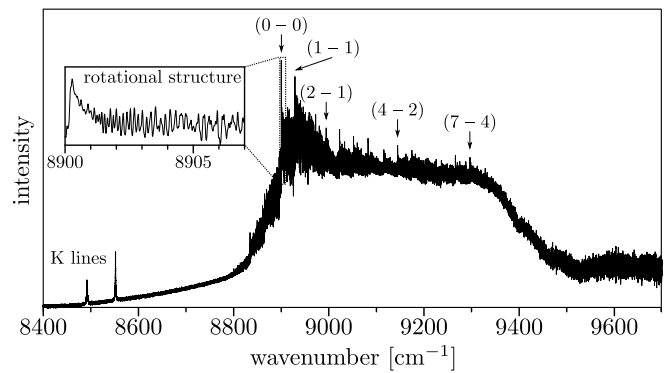}
	\caption[thermische Emission]{Record of the thermal emission of KCa. Rotational structure is shown enlarged by the inset. Some band heads of the vibrational structure are labeled.}
	\label{fig:emi}
\end{figure}

The thermal emission is observed in the range from \SI{8000}{\kay} to \SI{9700}{\kay}. \figref{fig:emi} shows the spectrum recorded with a resolution of 0.03 \si{\kay} and represents an average of 290 scans. On the left, one can clearly see lines at \SI{8491.8}{\kay} and \SI{8551.8}{\kay}, which correspond to the transitions $^2\mathrm{P}_{3/2}$ --  $^2\mathrm{D}_{5/2}$ and $^2\mathrm{P}_{1/2}$ -- $^2\mathrm{D}_{3/2}$ of atomic potassium. The KCa spectrum of the \bstate \  -- \Xstate \ band is located between \SI{8800}{\kay} and \SI{9400}{\kay}, as expected from the ab initio calculations \cite{pototschnig_vibronic_2017}. The shape of the background in this area changes with the temperature in the heatpipe. Since the conditions in the heatpipe change over time, the recordings have different background shapes but this does not influence the position of the lines.

One can well recognize band heads, the largest of which is at \SI{8900.3}{\kay} and is assigned to the \bands{0}{0} band on the basis of Franck-Condon factors (FCFs) derived from the ab initio potentials. The inset in \figref{fig:emi} is a zoomed part showing clearly the rotational structure.

Due to the atomic masses of about \SI{40}{\atomicmassunit} and the fairly shallow ground state potential, the rotational and vibrational constants of \ce{KCa} are rather small. Furthermore, the high temperature needed for producing the molecules leads to a wide range of populated vibrational and rotational levels and due to the Franck-Condon factors thermal emission is distributed over many vibrational transitions. The vibrational bands are strongly overlapping which hinders the direct assignment. Therefore, laser induced fluorescence (LIF) is used to identify pairs of lines according to the selection rule $\Delta J=\pm 1$ for the common total angular momentum of the excited state, $J\upS$, and associate lines of a vibrational progression within the ground state.

To excite the KCa molecules, a tunable diode laser with an antireflection coating (Toptica) is used in a Littrow configuration, and it is stabilized by a wavemeter (WS-U, HighFinesse GmbH), leading to an uncertainty of the exciting laser frequency of less than \SI{20}{\MHz}. The frequency range of the diode is about 8730 -- \SI{9350}{\kay}. The laser light with a power of 10-20 mW and a beam diameter of approximately \SI{2}{\mm} is precisely adjusted along the optical axis, which was fixed before by the fluorescence observation of K$_2$ induced by a HeNe-laser.  Lines selected from the emission spectrum are excited and the fluorescence spectra were recorded with a resolution of \SI{0.05}{\kay}. The fluorescence progression appears as enhanced lines in the thermal emission. An overview of the fluorescence studies is given in \tabref{tab:NRanges} (a).

Because a recorded spectral line consists of several overlapping Doppler-broadened transitions, we assign an uncertainty of \SI{0.02}{\kay} to most transition frequencies. For broad lines, a higher uncertainty is assumed. The measured frequency differences have been generally given an uncertainty of \SI{0.01}{\kay} if the fluorescence lines are sufficiently enhanced compared to the thermal emission spectrum because of the significantly reduced Doppler shift due to a mismatch of the laser excitation..

\section{Analysis}
\subsection{Overview}

\begin{figure}[h]
	\centering
	\includegraphics[width=\columnwidth]{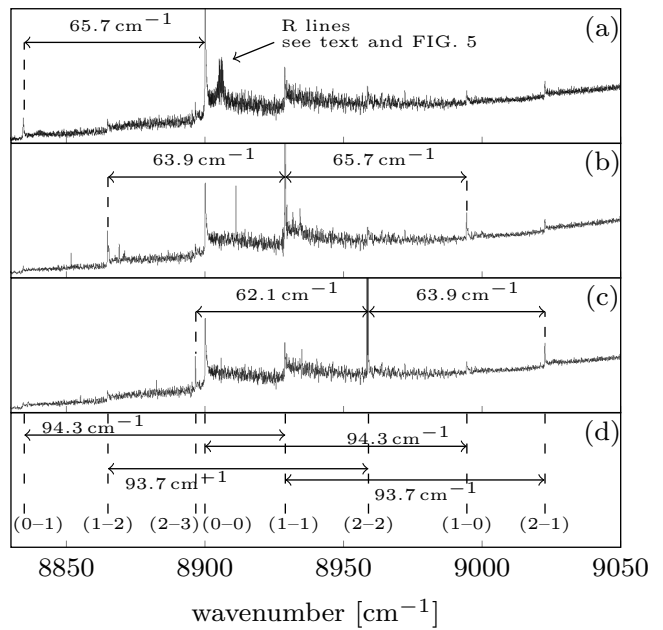}
	\caption{ Exemplary LIF spectra with excitations of different band heads show three linked systems of bands ((a) to (c)). Identifying equal distances between these bands allows for connecting these systems to a coherent structure of vibrational levels in the ground state. Selecting identical $v\loS$ gives the vibrational levels in the excited state (d).}
	\label{fig:spectra}
\end{figure}

Laser excitations in the band heads yield easily recognizable fluorescence (by the enhanced intensity compared to primary thermal emission) in other band heads which can be used to order the vibrational structure of the \Xstate \ and \bstate \  states. \figref{fig:spectra} (a) to (c) shows eight vibrational bands which are associated in three systems of almost equal vibrational spacing, representing the vibration of the ground state. The laser excitation is at the overshooting line. The short progressions also allowed for a preliminary assignment of vibrational quantum numbers starting with $v\loS=0$, since no fluorescence could be observed beyond this labeled line. This assignment is consistent with the later calculated FCFs (see section \ref{sec:rNd}). In \figref{fig:spectra}, the three progressions are shown above each other on the same absolute scale. The vibrational ladder for the excited state becomes visible by differences between two of such recordings as indicated at the bottom of the figure. The obtained vibrational spacings allowed for extrapolating the vibrational ladders. The band heads predicted in this way coincide with other band heads observed  up to \bands{7}{4} in the thermal emission spectrum, even if they showed no or hardly recognizable fluorescence, as indicated in \figref{fig:emi}.

\begin{figure}[h]
	\centering
	\includegraphics[width =\columnwidth]{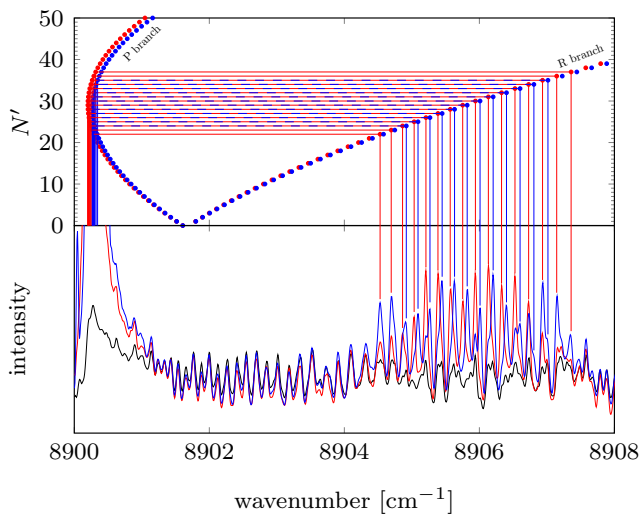}
	\caption{(Colour online) The red and blue spectra show LIF from excitations in the P band head of the \bands{0}{0} band of \ce{KCa} and fluorescence lines in the R branch. The thermal emission spectrum (black) is shown for reference. The upper part of the figure shows the Fortrat parabolas corresponding to the P and R branches for $N\upS$ separating the overlaps in the band head to the wide spread in the R branch. Red and blue circles correspond to \fone \ and \ftwo \ levels, respectively. The two cases of excitations show the differences between \fone \ and \ftwo: the red spectrum contains only florescence of \fone \ lines, the blue spectrum contains mainly fluorescence of \ftwo \ lines with some \fone \ lines at the edges of the fluorescence region. }
	\label{fig:parabola}
\end{figure}

\begin{figure}[h]
	\centering
	\includegraphics[width=\columnwidth]{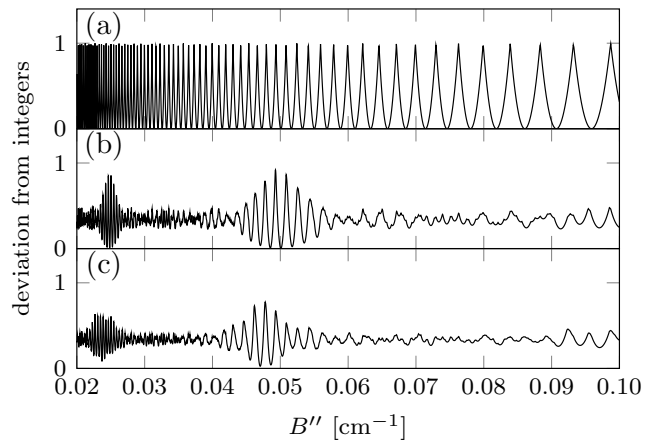}
	\caption{Accumulated deviation from integer numbers for $N\upS$ for (a) one $\Delta \nu$ measurement, (b) six $\Delta \nu$ measurements and (c) eleven $\Delta \nu$ measurements from a LIF experiment in the \bands{0}{0} band head. Notice that for (b) the global minimum does not give yet the correct value. For more than eleven values, the general shape of the curve is not altered.}
	\label{fig:BPlot}
\end{figure}

The bands are blue shaded, thus the band head is given by the P branch. Excitations in the band heads could yield fluorescence of a group of rotational lines in the R branch as seen in trace (a) of \figref{fig:spectra} above the band head \bands{0}{0}. These rotational fluorescences form two distinct systems, later identified as transitions between spin-up states or spin-down states, labeled \fone \ and \ftwo (see section \ref{subsec:dunham}). \figref{fig:parabola} depicts these fluorescences in detail, explaining the appearance of the group of rotational lines and the possibility to observe the energy difference between \fone \ and \ftwo \ lines by only slightly shifting the excitation frequency in the band head. We should point out that the experiment does not give information which line system is \fone \ or \ftwo. Similar groups of fluorescence lines were also observed for the \bands{0}{1} band.
The lines of such a group correspond to consecutive rotational quantum numbers $N$. (We use $N$ as rotational quantum number, see later Hund's coupling case(b)). For relatively small ranges of $N$, the changes in the effective rotational constant $B$ are negligible, thus the frequency differences $\Delta \nu$ of the PR pairs for $N\upS$ follow the formula
\begin{equation}
	\Delta \nu = B\loS \, (4\, N\upS + 2),
\label{eq:Dnu}
\end{equation}
Requiring that $N\upS$ is integer, a discrete set of values for the rotational constant is derived from each observed $\Delta \nu$.
For several observed PR differences, we construct the sum over the weighted quadratic distances to the closest integer for each observation with index $i$,
\begin{equation}
	\frac{1}{\#_{\mathrm{obs}} }\sum_i \left( \frac{ N\upS_i(B) - \mathcal{N}\upS_i(B) }{\delta \nu_i} \times 2 \, \delta \nu_{min} \right) ^2,
	\label{eq:sum}
\end{equation}
and the minimum of this function with respect to $B$ gives the value for $B\loS$ of the studied band. Here, $\#_{\mathrm{obs}}$ is the number of observations, $ \mathcal{N}\upS_i(B)$ is calculated with the observed $\Delta \nu_i$ from rearranging \equref{eq:Dnu}, $N\upS_i(B)$ is the integer closest to $\mathcal{N}\upS_i$, $\delta \nu_i$ is the experimental uncertainty of $\Delta \nu_i$ and the factor $2\cdot\delta \nu_{min}$, with the smallest uncertainty of the series, serves to normalize the expression. \figref{fig:BPlot} exemplifies such sums in dependence of a given rotational constant $B\loS$. Trace (a) is the example of a single observation and the zero-positions represent the discrete set of possible $B\loS$ mentioned above. Adding more and more observations (from (b) to (c)), the desired minimum of the function becomes more pronounced.
The value for $B\loS$ \ determined with this procedure from \figref{fig:BPlot} (c) is \SI{0.0470}{\kay} and close to the later determined value of $\mathrm{Y}\loS_{01}$ of the full evaluation (see \tabref{tab:Dunham}). 

Having thus determined $B \loS$, \equref{eq:Dnu} could directly be applied to assign rotational quantum numbers to the observed fluorescence lines and fluorescence lines in other vibrational bands from the same LIF spectrum. These assignments together with $B\loS$ can immediately be used to determine a value for $B\upS$. Thus LIF experiments with the laser tuned to several band heads provided the starting point for the analysis of the rovibrational spectrum.

\begin{table*}
	\caption{(a) Ranges of $N$ levels observed in each band via LIF experiments. (b) Ranges of all assigned $N$ levels.}
	\label{tab:NRanges}
	\begin{ruledtabular}
	\begin{tabular}{c*{17}c}
		\noalign{\vskip 5pt} 
	(a)	&                          &        &        &       & &\hspace{1cm}& (b)     &                &       &       &       &       &      &        &       &       &        \\
		&\diagbox{$v\loS$}{$v\upS$}& 0      & 1      & 2     & 3    &     & &\diagbox{$v\loS$}{$v\upS$}& 0     & 1     & 2     & 3     & 4     & 5     & 6     & 7     & 8      \\
		& 0                        & 8-162  & 46-145 &       &      &     & & 0                        &8-174  &46-176 &119-175&146-171&       &       &       &       &        \\
    	& 1                        & 4-162  & 14-123 & 9-132 & 62-93&     & & 1                        &4-162  &14-123 &9-150  &62-160 &125-144&       &       &       &        \\
		& 2                        &131-155 & 34-143 & 9-68  & 62-93&     & & 2                        &131-155&34-143 &9-68   &39-103 &44-147 &49-149 &115-147&       &        \\
		& 3                        &        & 84-145 & 9-103 &      &     & & 3                        &       &84-146 &9-110  &       &47-80  &32-108 &48-133 &       &        \\
		& 4                        &        &        &54-123 & 62-93&     & & 4                        &       &137-154&21-123 &62-93  &       &       &19-79  &41-85  &        \\
		& 5                        &        &        &119-132& 62-93&     & & 5                        &       &       &119-132&47-93  &       &77-84  &       &       &39-89   \\
	\end{tabular}
	\end{ruledtabular}
	
\end{table*}

\subsection{Assignment and molecular parameters}
\label{subsec:dunham}
\begin{figure}
	\centering
	\includegraphics[width=1.0\columnwidth]{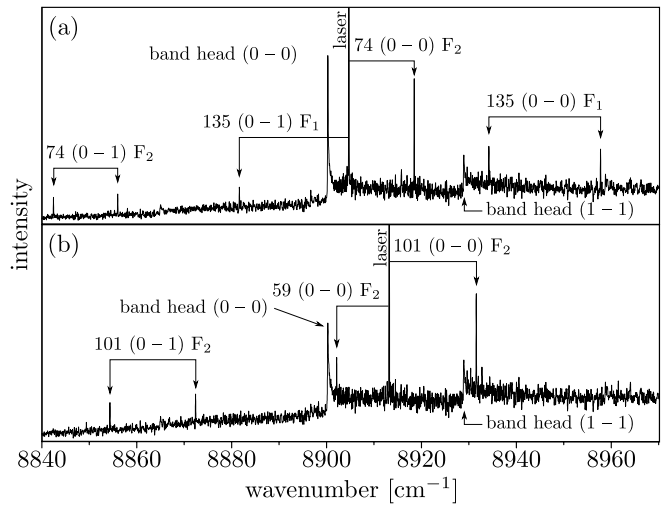}
	\caption[LIF]{Laser induced fluorescence of two excitations at \SI{8904.7115}{\kay} (a) and \SI{8913.2055}{\kay} (b). The P-R pairs are labeled with the rotational quantum number $N^\prime$ of the excited state, vibrational band and the component of the spin-rotation coupling.}
	\label{fig:lif}
\end{figure}
Hund's coupling case (b) with the basis vector $\ket{\Lambda, (N,S)J}$ is used to describe both $^2\Sigma^+$ states, where $\Lambda$ is the quantum number associated with the projection of the orbital angular momentum to the molecular axis (here $\Lambda=0$), $\hat{N}$ is the total angular momentum without spins, $\hat{S}$ is the total spin of the electrons and $\hat{J} = \hat{N} + \hat{S}$ is the total angular momentum of the molecule excluding the nuclear spins. The rovibrational energies can be expressed by the Dunham expansion
\begin{equation}
E(v,N) = \sum_{m,n} \mathrm{Y}_{m n}(v+\frac{1}{2})^m[N(N+1)]^n,
\label{eq:DunhamExp}
\end{equation}
for each electronic state with the so-called Dunham parameters Y$_{m n}$. The parameter $\mathrm{Y_{00}}$ is used in this analysis to approximate the electronic term energy $T_e$ and thus has a different meaning than the conventional Dunham correction $\mathrm{Y_{00}}$.

The energy levels with $J=N+1/2$ and $J=N-1/2$ of a doublet state $(S=1/2)$ are labeled by \fone \ and \ftwo, respectively, and are split for the same $N$ because of the spin-rotation coupling given by the Hamiltonian $\gamma\hat{S}\cdot\hat{N}$ with a coupling constant $\gamma$\cite{herzberg_spectra_1950}. This energy is added to the rovibrational energies since the interaction operator is diagonal in the Hund's case (b) basis:
\begin{subequations}
	\begin{align}
	E&= E(v,N) + \frac{\gamma}{2} \times N  &\quad\mathrm{for\; F}_1 \\
	E&= E(v,N) - \frac{\gamma}{2} \times (N+1) &\quad\mathrm{for\; F}_2
	\end{align}
	\label{eq:SRcoupling}
\end{subequations}
In principle, the spin-rotation contribution will depend on the vibrational and rotational level and thus could be described by a corresponding Dunham-like series with $\gamma_{m n}$ as expansion parameters. Our evaluation program contains this form for both states, but the energy expression being linear in $N$ results for P and R transitions to a function which mainly depends on the differences of the spin-rotation interactions of the two electronic states. Thus we choose $\gamma_{0 0}$ of the excited state as non-zero and this value actually represents the difference of the two interactions. Because we have sets of fluorescence progressions, each with a common excited level, we found that the data contain a slight $\gamma$ dependence of the vibrational level of the ground state, thus we introduced for this fact the parameter $\gamma_{1 0}$ for the ground state. All other expansion parameters $\gamma_{mn}$ were set to zero.

The parameters $\mathrm{Y_{00}}$, $\mathrm{Y_{01}}$ for both states and $\gamma_{0 0}$ of the excited state can be fitted using the fluorescence spectra of the \bands{0}{0} band head excitations. Starting with these initial parameters, more fluorescence lines with assigned quantum numbers are subsequently included in the fit based on their PR differences. Similarly, fluorescence lines from other vibrational bands are included.

Since the KCa spectrum is dense, the laser we use in an LIF experiment can affect several transitions from different bands. Examples of such cases of fluorescence are given in \figref{fig:lif}. In this figure, two laser excitations are shown on the same absolute scale to illustrate the relation of the different vibrational bands. In both cases, two progressions are excited simultaneously. Trace (b) shows very different $N$ and results in Stokes pairs of the band \bands{0}{1} and trace (a) is an excitation of a line in \bands{0}{0} and one in \bands{0}{1} showing anti-Stokes fluorescence to \bands{0}{0}. 
\begin{table*}
	\caption{\label{tab:DunPar}Dunham and spin-rotation parameters for the first two $^2\Sigma^{+}$ states of \KCa. The parameters give an accurate description for levels with quantum numbers as given in \tabref{tab:NRanges}. All values given in \si{\kay}.}
	\label{tab:Dunham}
	\begin{ruledtabular}
		\begin{tabular}{rlllll}
			\noalign{\vskip 5pt} 
			\multicolumn{6}{c}{\Xstate} \\
		  $n$\phantom{$^2$}& $\mathrm{Y}_{n0}$           & $\mathrm{Y}_{n1}$            & $\mathrm{Y}_{n2}$          & $\mathrm{Y}_{n3}$        & $\mathrm{Y}_{n4}$  \\
			0\phantom{$^2$}& \num{0.0}                   & \num{4.75384 +- 0.00025	 e-2} & \num{-9.3968 +- 0.019 e-08} & \num{-8.54 +- 0.88 e-14} & \num{-6.57 +- 0.15 e-18} \\
			1\phantom{$^2$}& \num{67.9826 +- 0.0037 e-0} & \num{-6.4306 +- 0.0084 e-4}  & \num{-2.062 +- 0.064 e-9}  & \num{-7.20 +- 0.14 e-14} & \multicolumn{1}{c}{-} \\
			2\phantom{$^2$}& \num{-9.3950 +- 0.0045 e-1} & \num{-5.87 +- 0.15 e-6}      & \num{-1.771 +- 0.074 e-10} & \multicolumn{1}{c}{-}    & \multicolumn{1}{c}{-} \\
			3\phantom{$^2$}& \num{-1.01 +- 0.15 e-3} & \multicolumn{1}{c}{-}      & \multicolumn{1}{c}{-} & \multicolumn{1}{c}{-}    & \multicolumn{1}{c}{-} \\
$\gamma_{10}$\phantom{$^2$}& \num{2.05 +- 0.66 e-5}      &                              &                            &                          &   \\
			\hline
			\noalign{\vskip 5pt} 
			\multicolumn{6}{c}{\bstate} \\
		  $n$\phantom{$^2$}& $\mathrm{Y}_{n0}$           & $\mathrm{Y}_{n1}$            & $\mathrm{Y}_{n2}$           & \multicolumn{2}{c}{} \\
			0\phantom{$^2$}& \num{8888.0459 +- 0.0036 }  & \num{4.88331 +- 0.00022 e-2} &\num{-5.0278 +- 0.0075 e-8}  & \multicolumn{2}{c}{} \\
			1\phantom{$^2$}& \num{94.9595 +- 0.0029}     & \num{-1.9686 +- 0.0031 e-4}  &\num{-1.446 +- 0.089 e-10}   & \multicolumn{2}{c}{} \\
			2\phantom{$^2$}& \num{-2.7120 +- 0.0085 e-1 }& \num{-1.049 +-0.031 e-6}    & \multicolumn{1}{c}{-}       & \multicolumn{2}{c}{} \\
			3\phantom{$^2$}& \num{-1.636 +- 0.074 e-3}   & \multicolumn{1}{c}{-}        & \multicolumn{1}{c}{-}       & \multicolumn{2}{c}{} \\
$\gamma_{00}$\footnote{Note that the experimental data only yields the difference, namely $\gamma_{00}\upS-\gamma_{00}\loS$.} & \num{-2.234 +- 0.014 e-3} &                              &                             & \multicolumn{2}{c}{} \\
		\end{tabular}
	\end{ruledtabular}
\end{table*}
From the fluorescence spectra, lines from bands with $v\upS=0-2$ and $v\loS=0-3$ could initially be identified and assigned. Reviewing the fluorescence spectra with the final molecular parameters yielded the identification of additional fluorescence lines with $v\upS = 3$ and $v\loS = 4$ -- 5. The ranges of observed rotational quantum numbers are listed in \tabref{tab:NRanges}.

Further bands are analyzed using the thermal emission spectrum.
For this purpose, the thermal emission spectrum is simulated with the Dunham parameters determined up to the actual intermediate evaluation step with Franck-Condon factors derived from the initial ab initio potentials and a line width estimated from the Doppler width and the selected resolution of the FTS and a population according to the actual temperature. This theoretical spectrum is compared with the recorded spectrum and allows for adding additional lines to the fit and to adjust the parameters and add new ones.
Occasionally, the simulated spectrum shows a high intensity line close to lines with similar intensity and width in the experimentally recorded spectrum. If these lines show this behavior for a long series of quantum numbers, they are identified with the lines from the recorded spectrum and a fit will show the consistency with the other assignments done before.
The extension in quantum number space was done in small steps, avoiding large extrapolation in quantum numbers $v$ and $N$, because due to the dense spectra it is very probable to assign a wrong line. By doing this carefully, the shape of the simulated spectrum converges to the shape of the experimental spectrum.

In total, 2554 lines and frequency differences could be measured between 878 levels in the ground state up to $v\loS=5$ and 549 levels in the excited state up to {$v\upS=8$}.
The resulting Dunham parameters are listed in \tabref{tab:Dunham} with their estimated standard deviations.
The normalized standard deviation $\sigma=0.95$ of this linear fit validates the choice of the uncertainties described in section \ref{sec:experiment}.
All assigned lines used in the the final fit of the Dunham parameters are contained in the supplementary material.

\section{Potential Fit}
The evaluation with the Dunham expansion shows that the two electronic states are well described by Hund's coupling case (b) with basis vectors $\ket{\Lambda,(N,S)J}$ and the possible coupling to other electronic states for example by spin-orbit coupling is not observed within the present experimental accuracy of \SI{0.02}{\kay}. Thus we can set up two separate Hamiltonians for the two states. Because the spin-rotation interaction is diagonal in the basis, the corresponding energy contribution adds to the rovibrational eigenenergies $E(v,N)$ as given in \equref{eq:SRcoupling}. The Schr\"odinger equation for the rovibrational energies and for the rovibrational wavefunction $\psi_{vN}$ has the conventional form
\begin{equation}
\begin{split}
 \left[-\frac{\hbar^2}{2m}\;
\frac{\mathrm{d}^2}{\mathrm{d}R^2}+U(R)+\frac{\hbar^2}{2mR^2}N(N+1)\right]
\psi_{vN}(R) \\
 =E(v,N)\psi_{vN}(R) \mbox{\ ,}
\end{split}
 \label{rad}
 \end{equation}
with the reduced mass $m$ of the molecule and the potential function $U(R)$ for each electronic state. The approach by potential functions gives the advantage to release the constraints by the truncation of the power expansion in quantum numbers $(v,N)$ for the energies in Dunham representation. Such possible effect can be contained in the solution in \tabref{tab:Dunham} looking at the different centrifugal contributions for the two electronic states, namely the highest term $\mathrm{Y}_{02}$ for the excited state and $\mathrm{Y}_{04}$ for the ground state.

For solving the Schr\"odinger equation, we set up the potential in the analytic form as successfully applied in several earlier papers (e.g. \cite{samuelis_cold_2000})
\begin{align}
U(R)&=\sum_{i=0}^{n}a_i\,\xi(R)^i, \\
\label{eq:xv}
\xi(R)&=\frac{R - R_m}{R + b\,R_m},
\end{align}

\noindent where the coefficients $a_i$ are fitted
parameters and $b$ and $R_m$ are fixed. $R_m$ is normally set close to the value of the
equilibrium separation $R_e$. The potential is extrapolated for $R <
R_{\rm inn}$ with:

\begin{equation}
\label{eq:rep}
  U(R)= A + B/R^{N_s}
\end{equation}

\noindent by adjusting the parameters $A$, $B$ with $N_s=6$ to get a continuous
transition at $R_{\rm inn}$.
For large internuclear separations ($R > R_{\rm out}$)
we adopted the standard long range form:
\begin{equation}
U_{\mathrm{LR}}=U_{\infty}-C_6/R^6-C_8/R^8-C_{10}/R^{10}
\label{lr}
 \end{equation}
The dispersion parameters $C_i$ do not play any role in our present case, because we observed only low vibrational levels and did not reach the long range region in $R$, thus these parameters serve for setting up a smooth potential function to the dissociation limit $U_{\infty}$. 

A non-linear least squares fit to all assigned transitions and transition differences with common upper levels obtained from the fluorescence study was performed, fitting simultaneously the potential functions for the two electronic states. The spin-rotation contribution was included in exactly the same form as used before, see  \equref{eq:SRcoupling}. The fit describes the experimental data with the same quality, namely the normalized standard deviation $\sigma=0.94$, as obtained for the Dunham fit. The derived potential parameters are given in TABLES \ref{tab_X} and \ref{tab_A}, the number of fit parameters is only 20, thus smaller than in the Dunham case (22), which demonstrates the compact form of the representation by potentials. If one looks close to the individual deviations between Dunham and potential approach, one sees differences mainly for high rotational levels N, which certainly is related to the above mentioned constraint by the truncation in the Dunham model. On average the deviations of both approaches are the same. An example of the systematic trend with quantum number $N$ between both approaches is included in the supplementary material.

\begin{table}[t]
\fontsize{8pt}{8pt}\selectfont 
\caption{Parameters of the
analytic representation of the potential for state \Xstate. Parameters with an asterisk $^\ast$ ensure smooth continuous extrapolation of the potential at $R_{\mathrm{inn}}$ and $R_{\mathrm{out}}$, with $^{\ast\ast}$ for constructing the dissociation asymptote.} 
\label{tab_X} 
\vspace{2mm}
	\begin{ruledtabular}
		\begin{tabular*}{1.0\linewidth}{@{\extracolsep{\fill}}lr}
		\multicolumn{2}{c}{$R < R_\mathrm{inn}= 3.72$ \AA} \\
		\hline
		$A^\ast$ & $-0.17137691\times 10^{4}$ \si{\kay} \\
		$B^\ast$ & $ 0.623058968\times 10^{7}$ \si{\kay} \AA $^{6}$ \\
		\hline\vspace{.3pt}\\
		\multicolumn{2}{c}{$R_\mathrm{inn} \leq R \leq R_\mathrm{out}=  5.3$ \AA} \\
		\hline
		$b$ & $-0.50$ \\
		$R_\mathrm{m}$ & $4.23968785$ \AA \\
		 $a_{0}$ & $ 0.000010$ \si{\kay}\\
		 $a_{1}$ & $ 0.916715672820951011$   \si{\kay}\\
		 $a_{2}$ & $ 0.607228674244838840\times 10^{ 4}$ \si{\kay}\\
		 $a_{3}$ & $-0.661911115663710007\times 10^{ 3}$ \si{\kay}\\
		 $a_{4}$ & $-0.610926466028853429\times 10^{ 4}$ \si{\kay}\\
		 $a_{5}$ & $-0.180355143531396743\times 10^{ 4}$ \si{\kay}\\
		 $a_{6}$ & $-0.262462209324614050\times 10^{ 5}$ \si{\kay}\\
		 $a_{7}$ & $ 0.942565374511957998\times 10^{ 4}$ \si{\kay}\\
		 $a_{8}$ & $ 0.360954584572881111\times 10^{ 6}$ \si{\kay}\\
		 $a_{9}$ & $-0.637614449251196464\times 10^{ 6}$ \si{\kay}\\
		\hline\vspace{.3pt}\\
		\multicolumn{2}{c}{$R_\mathrm{out} < R$}\\
		\hline
		${U_\infty^{\ast\ast}}$ & 1475.0 \si{\kay} \\
		${C_6}$\cite{mitroy03} & 0.13508721$\times 10^{8}$ \si{\kay}\AA$^6$ \\
		${C_{8}^{\ast}}$ & 0.11594031$\times 10^{10}$ \si{\kay}\AA$^8$ \\
		${C_{10}^\ast}$ & -0.27203341$\times 10^{11}$ \si{\kay}\AA$^{10}$ \\
		\hline\vspace{.3pt}\\
		\multicolumn{2}{c}{for all R}\\
		\hline
		$\gamma_{10}$ & $0.228\times 10^{-4}$ \si{\kay}\\
		\end{tabular*}
	\end{ruledtabular}
\end{table}
%
%
\begin{table}[h]
\fontsize{8pt}{8pt}\selectfont 
\caption{Parameters of the
analytic representation of the potential for state \bstate. Parameters with an asterisk $^\ast$ ensure smooth continuous extrapolation of the potential at $R_{\mathrm{inn}}$ and $R_{\mathrm{out}}$, with $^{\ast\ast}$ for constructing the dissociation asymptote.} 
\label{tab_A}
\vspace{2mm}

	\begin{ruledtabular}
		\begin{tabular*}{1.0\linewidth}{@{\extracolsep{\fill}}lr}

		\multicolumn{2}{c}{$R < R_\mathrm{inn}= 3.605$ \AA} \\
		\hline
		$A^\ast$ & $0.67975020\times 10^{4}$ \si{\kay} \\
		$B^\ast$ & $0.73133965\times 10^{7}$ \si{\kay} \AA $^{6}$ \\
		\hline\vspace{.3pt}\\
		\multicolumn{2}{c}{$R_\mathrm{inn} \leq R \leq R_\mathrm{out}=  5.1$ \AA} \\
		\hline
		$b$ & $-0.24$ \\
		$R_\mathrm{m}$ & $4.18303794$ \AA \\
		 $a_{0}$ & $8888.0195 $\si{\kay}\\
		 $a_{1}$ & $ 8.97211680301065329$   \si{\kay}\\
		 $a_{2}$ & $ 0.266368790864483090\times 10^{ 5}$ \si{\kay}\\
		 $a_{3}$ & $ 0.672648244268816416\times 10^{ 4}$ \si{\kay}\\
		 $a_{4}$ & $-0.112400340636989258\times 10^{ 5}$ \si{\kay}\\
		 $a_{5}$ & $-0.904918359031159489\times 10^{ 5}$ \si{\kay}\\
		 $a_{6}$ & $-0.360117376114208950\times 10^{ 6}$ \si{\kay}\\
		 $a_{7}$ & $ 0.296505843717056094\times 10^{ 7}$ \si{\kay}\\
		 $a_{8}$ & $ 0.530811059962429386\times 10^{ 7}$ \si{\kay}\\
		 $a_{9}$ & $-0.538588415900813341\times 10^{ 8}$ \si{\kay}\\
		\hline\vspace{.3pt}\\
		\multicolumn{2}{c}{$R_\mathrm{out} < R$}\\
		\hline
		${U_\infty^{\ast\ast}}$ & 14500.0 \si{\kay} \\
		${C_6}^{\ast\ast}$ & 0.10$\times 10^{8}$ \si{\kay}\AA$^6$ \\
		${C_{8}^{\ast}}$ & 0.68834939$\times 10^{10}$ \si{\kay}\AA$^8$ \\
		${C_{10}^\ast}$ & -0.13481830$\times 10^{12}$ \si{\kay}\AA$^{10}$ \\
		\hline\vspace{.3pt}\\
		\multicolumn{2}{c}{for all R}\\
		\hline
		$\gamma_{00}$\footnote{Note that the experimental data only yields the difference, namely $\gamma_{00}\upS-\gamma_{00}\loS$.}  & -0.0022317 \si{\kay}  \\
		\end{tabular*}
	\end{ruledtabular}
\end{table}

The potential functions derived by the potential fit are overlaid in \figref{fig:PotentialCurves} by thick lines in the region where they are determined form the observations. The agreement looks nice, but the differences hidden by the thick lines are in the order of 100 \si{\kay}.

The potentials are also used to extend iteratively the assignment of the emission spectra, comparing the simulation of the spectra constructed by energies and FCFs with the thermal emission. We also calculated the spectrum of the second most abundant isotope combination of KCa, namely \ce{^{41}K^{40}Ca} with \SI{6.5}{\percent} natural abundance, by mass scaling. At several places, the intensity of such lines would be large enough for observation but the overlap with other lines is too strong to obtain an unambiguous assignment.
 
\section{Discussion and Outlook}
\label{sec:rNd}

The fluorescence lines in the laser-addressed region (8820 -- \SI{9050}{\kay}) are clearly visible up to $v\upS=2$ and $v\loS=3$. The systematic comparison of the recorded spectrum of the thermal emission with the simulation applying Dunham parameters or potentials yields the description up to $v\upS=8$ and $v\loS=5$. Overall, lines in the range 8820 -- \SI{9380}{\kay} are assigned. The FCFs generated with the potentials (TABLES \ref{tab_X} and \ref{tab_A}) show for $N=0$ in the range $v\upS=0$ -- 8 and $v\loS=0$ -- 8 a structure in magnitude similar to that in the ab initio work in reference \cite{pototschnig_vibronic_2017}, thus their conclusion about creation of could molecules stays valid. In addition, the FCFs show a strong rotational dependency. When increasing $N$ to large values, the distribution of significant FCFs covers a larger interval of vibrational levels (see FCF tables in the supplement).

\begin{figure}
  	\centering
  	\includegraphics[width=1.0\columnwidth]{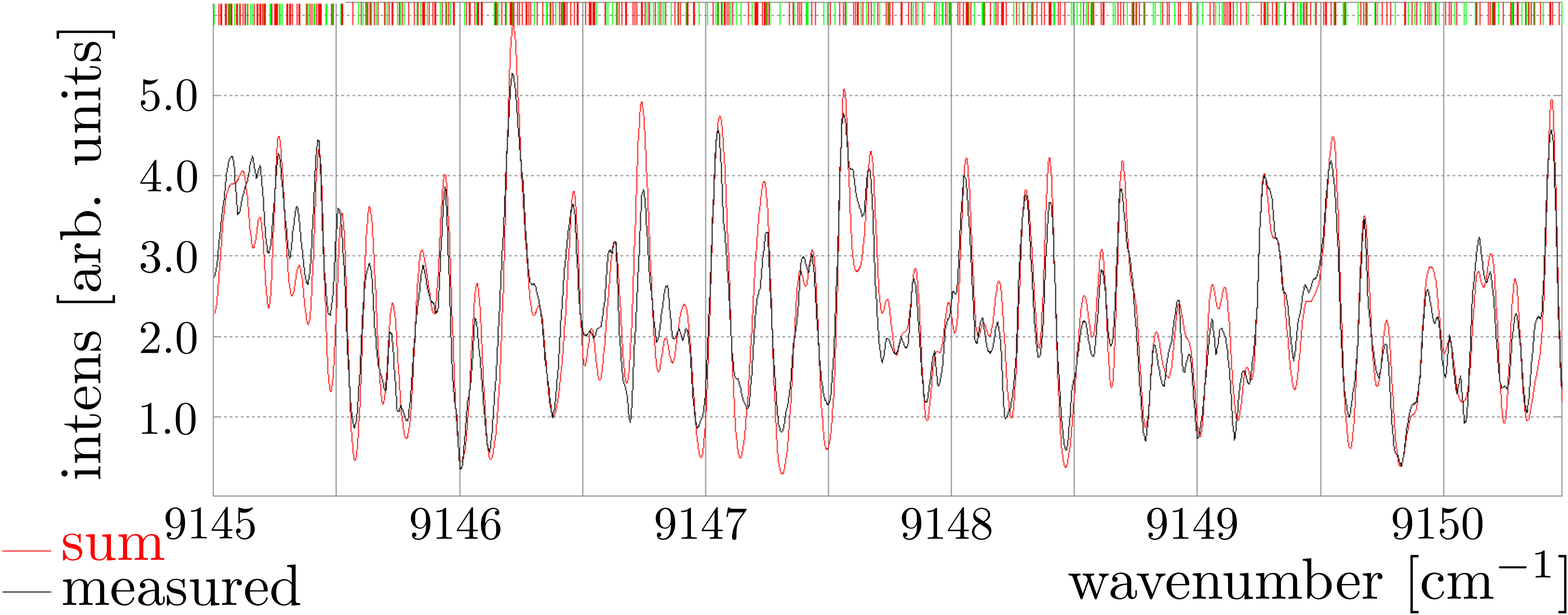}
  	\caption{(Colour online) Simulation of the spectrum in the region of \num{9145}-- \SI{9150.5}{\kay}. The black curve represents the experimental spectrum, the red curve represents the sum of the simulated spectra for the isotopologues \ce{^{39}K^{40}Ca} (\SI{90.4}{\percent} natural abundance) and \ce{^{41}K^{40}Ca} (\SI{6.5}{\percent} natural abundance). The red (\ce{^{39}K^{40}Ca}) and green (\ce{^{41}K^{40}Ca}) bars in the upper part of the diagram represent calculated transitions for both isotopologues. For convenience, we have edited the text for better readability.}
  	\label{fig:sim}
\end{figure}

The simulation of the thermal emission spectrum with the potentials from TABLES \ref{tab_X} and \ref{tab_A} reproduces the recorded spectrum in the range 8835 -- \SI{9340}{\kay} well. High consistency is found in the regions 8900 -- \SI{8940}{\kay}, dominated by the \bands{0}{0} and \bands{1}{1} bands, 9040 -- \SI{9085}{\kay}, dominated by the \bands{1}{0}, \bands{2}{1} and \bands{3}{2} bands and 9120 -- \SI{9190}{\kay}, dominated by the \bands{2}{0}, \bands{3}{1}, \bands{4}{2} and \bands{5}{3} bands. For these bands, we have many assigned lines (see \tabref{tab:NRanges} (b)). The spectral range 8940 -- \SI{9010}{\kay} is less well described. This is mainly due to the fact that in this region lie lines of bands with $v\loS=6$ -- 10 and good FCFs that were not taken into account in the potential fit. Due to the strong overlap of the bands, even small displacements of these lines by incomplete modeling strongly influence the overall intensity profile. The spectrum below \SI{8820}{\kay} and beyond \SI{9380}{\kay} cannot be analyzed convincingly because the intensity is too low (see \figref{fig:emi}).
A screenshot from the simulation program is shown in \figref{fig:sim}. In the upper part the individual transition positions are indicated by colored vertical bars. Notice that small deviations of the measured and simulated profiles appear throughout the diagram. They can mostly be attributed to result from the sum of many lines, each might deviate within the uncertainty of the model from the ``true'' transition frequency, see the dense pattern in the upper part of the figure. Lines for the second abundant isotopologue \ce{^{41}K^{40}Ca} are extrapolated since all transitions could be attributed to the main isotopologue \ce{^{39}K^{40}Ca}.
With the help of the simulation, we revisited the fluorescence studies. The systematic inspection shows relatively long vibrational progressions with fluorescence intensities in agreement with the predicted FCFs. Many such fluorescence lines were not sufficiently pronounced compared to the thermal emission, so that they were overlooked in the former assignment process. This successful extension shows the overall consistency of the whole simulation despite the existence of less well described spectral regions mentioned above.

	\begin{table*}
		\caption{\label{tab:constComp}Comparison of measured spectroscopic constants of \KCa with results of known ab initio calculations. All values are given in \si{\kay}, except $R_e$ which is given in \AA. }
		\begin{ruledtabular}
			\begin{tabular}{clccccccr}
				&Method&$R_e$ &$D_e$ &$\omega_e \approx \mathrm{Y}_{10}$&$\omega_e x_e \approx -\mathrm{Y}_{20}$ &$B_e \approx \mathrm{Y}_{01}$ &$T_e $ & Ref.\\
				\hline
				\noalign{\vskip 5pt} 
				\multirow{3}{*}{\Xstate}
				&CCSD(T)       &4.32\phantom{00}&\phantom{0}974&61\phantom{.000}&-               &0.045\phantom{000}&0 & \cite{gopakumar_dipole_2014}\\
				&MRCI          &4.197\phantom{0}&1474          &70.8\phantom{00}&0.85\phantom{0}&-                 &0 & \cite{pototschnig_vibronic_2017}\\
				&Dunham        &4.2377          &-             &67.983          &0.940           &0.047538          &0 & this work\\
				&potential fit &4.2395          &-             &67.971          &-               &0.047541          &0 &this work\\
				\hline
				\noalign{\vskip 5pt}    
				\multirow{2}{*}{\bstate}
				&MRCI          &4.177\phantom{0}   & 5633 & 95.0\phantom{00}& 0.40\phantom{0}& -      & 8922\footnote{derived from potential given in private communications}\phantom{.00} & \cite{pototschnig_vibronic_2017}\\
				&Dunham        &4.1811             & -    & 94.960          &0.271             &0.048831&8888.047& this work\\
				&potential fit &4.1825             & -    & 94.937          &-                 &0.048846&8888.019& this work\\
			\end{tabular}
		\end{ruledtabular}
	\end{table*}
	
Both evaluation approaches, the energy representation by Dunham coefficients or by molecular potentials, fit the measurement data equally well. \tabref{tab:constComp} shows for comparison the molecular parameters determined in this work and those from known ab~initio calculations. The agreement of the fitted parameters is very good and differences between Y$_{10}$ defined by the Dunham approach and $\omega_e$ for the potentials are not significant but they are different in definition and this is also true for Y$_{01}$ and $B_e$. The experimentally determined equilibrium separation $R_e$ and vibrational constant $\omega_e$ of the ground state lie between the two ab initio values. In the excited state, these theoretical values deviate from the experimental ones by less than \SI{1}{\percent}. The potential depth could not be derived because of missing high vibrational states. The experimentally determined electronic term energy of the excited state differs only by \SI{34}{\kay} from the calculated value, confirming the good quality of the ab initio calculations. 

Compared to the lighter molecule LiCa\cite{stein_spectroscopic_2013}, the rotational and spin rotational constants of KCa are smaller. KCa does not show any perturbations of the $2 ^2\Sigma^+$ state within the achieved accuracy, in contrast to LiSr where we observed the coupling between \bstate and \astate \ in our previous work\cite{LiSrFTS}.
 
Because of the high density of the lines in the KCa emission spectrum, the LIF experiments were indispensable for assignment. To extend these results to higher vibrational bands, we will start a new series of laser excitations. We expect to detect lines up to $v\loS=10$ of the ground state because of sufficiently large FCFs. For further investigations of KCa, we will study $(3) ^2\Sigma^+$~--~\Xstate \ transitions in the visible spectral range for which a high transition dipole moment is predicted\cite{pototschnig_vibronic_2017}. The relevant spectral region around \SI{14000}{\kay} should be sufficiently devoid of \ce{K_2} spectra\cite{johnson_continua_1985}, whereas \ce{Ca_2} lines can be expected \cite{allard_study_2005}, but should not dominate the spectrum. These proposed studies will provide higher vibrational levels of the ground state needed for extrapolating to the atom pair asymptote which is greatly desired for the study of ultracold KCa.\\

\section*{supplementary material}

	See the supplementary material for the full recorded thermal emission spectrum, a list of the assigned lines, an overview of the deviations of the Dunham and potential model and derived tables of Franck-Condon factors for various $N$.

\section*{acknowledgments}
	We thank Alexander Stein for his contribution in the initial phase of this project.

	This work received financial support from the Deutsche Forschungsgemeinschaft (DFG).


\section*{references}
\bibliography{KCa_17}

\end{document}